%% file: acl_latex.tex
\pdfoutput=1

\documentclass[11pt]{article}

\usepackage[final]{acl}

\usepackage{times}
\usepackage{latexsym}

\usepackage[T1]{fontenc}

\usepackage[utf8]{inputenc}

\usepackage{microtype}

\usepackage{inconsolata}

\usepackage{graphicx}

\input{config.tex}

\title{DecompileBench: A Comprehensive Benchmark for Evaluating Decompilers in Real-World Scenarios}

\author{
  Zeyu Gao\textsuperscript{1}\thanks{Equal contributions},
  Yuxin Cui\textsuperscript{1}\footnotemark[1],
  Hao Wang\textsuperscript{1}\footnotemark[1],
  Siliang Qin\textsuperscript{2}\footnotemark[1] \\
  \textbf{Yuanda Wang\textsuperscript{3},
  Zhang Bolun\textsuperscript{2},
  Chao Zhang\textsuperscript{1}\thanks{Corresponding author}} \\
  \textsuperscript{1}Tsinghua University \textsuperscript{3}Peking University \\
  \textsuperscript{2}Institute of Information Engineering, Chinese Academy of Sciences \\
  \texttt{\{gaozy22,yx-cui24,hao-wang20\}@mails.tsinghua.edu.cn} \\
  \texttt{\{qinsiliang,zhangbolun\}@iie.ac.cn} \quad
  \texttt{yuandawang958@stu.pku.edu.cn} \\
  \texttt{chaoz@tsinghua.edu.cn} \\
}

\begin{document}

\maketitle
\begin{abstract}
\input{abstract.tex}
\end{abstract}

\balance

\input{body.tex}

\balance

\clearpage
\newpage

\bibliography{elsa}

\clearpage
\newpage

\appendix

\input{appendix.tex}

\balance

\end{document}

%% file: config.tex
\usepackage{verbatim}
\usepackage{amsmath}

\usepackage{amssymb}
\usepackage{algorithmic}
\usepackage{algorithm}
\usepackage{makecell}
\usepackage{boxedminipage}
\usepackage{svg}
\usepackage{amsmath}
\definecolor{shadecolor}{RGB}{246,246,246}

\usepackage{graphicx}
\usepackage{epstopdf}
\usepackage{array}
\usepackage{booktabs}
\usepackage{multirow}
\usepackage{colortbl}
\usepackage{caption} 
\usepackage{multicol}
\usepackage{tabularx}
\usepackage{framed}
\usepackage{enumitem}
\usepackage{threeparttable}
\usepackage{supertabular}

\graphicspath{{figs/}}

\usepackage{colortbl} 
\definecolor{jade}{rgb}{0.0, 0.66, 0.42}
\definecolor{carolinablue}{rgb}{0.6, 0.73, 0.89}
\definecolor{dkgreen}{rgb}{0,0.6,0}
\definecolor{dkblue}{rgb}{0,0.4,0.5}
\definecolor{gray}{rgb}{0.5,0.5,0.5}
\definecolor{mauve}{rgb}{0.58,0,0.82}

\definecolor{codegreen}{rgb}{0,0.6,0}
\definecolor{codegray}{rgb}{0.5,0.5,0.5}
\definecolor{codepurple}{rgb}{0.58,0,0.82}
\definecolor{codeblue}{rgb}{0,0,205}
\definecolor{backcolour}{rgb}{245,245,245}

\usepackage{listings}
\usepackage{enumitem}

\lstdefinelanguage
   [x64]{Assembler}     
   [x86masm]{Assembler} 
   {morekeywords={xend, CDQE,CQO,CMPSQ,CMPXCHG16B,JRCXZ,LODSQ,MOVSXD, %
                  POPFQ,PUSHFQ,SCASQ,STOSQ,IRETQ,RDTSCP,SWAPGS, %
                  rax,rdx,rcx,rbx,rsi,rdi,rsp,rbp, %
                  r8,r8d,r8w,r8b,r9,r9d,r9w,r9b, %
                  r10,r10d,r10w,r10b,r11,r11d,r11w,r11b, %
                  r12,r12d,r12w,r12b,r13,r13d,r13w,r13b, %
                  r14,r14d,r14w,r14b,r15,r15d,r15w,r15b}} 

\lstdefinestyle{mystyle}{
    backgroundcolor=\color{backcolour},   
    commentstyle=\color{codegreen},
    keywordstyle=\color{codeblue},
    numberstyle=\tiny\color{codegray},
    stringstyle=\color{codeblue},
    basicstyle=\ttfamily\footnotesize,
    breakatwhitespace=false,         
    breaklines=true,                 
    captionpos=b,                    
    keepspaces=true,                 
    numbers=left,                    
    numbersep=5pt,                  
    showspaces=false,                
    showstringspaces=false,
    showtabs=false,                  
    tabsize=2
}
\usepackage{url}

\usepackage{flushend}
\usepackage{balance}
\usepackage{natbib}

\usepackage[misc]{ifsym}
\usepackage{bbding}

\usepackage{xspace}

\newcommand{\sysname}{{\tt DecompileBench}\xspace}

\usepackage{xspace}
\usepackage{subfig}

\usepackage{tabularx}
\usepackage{ragged2e}
\usepackage{booktabs}
\usepackage{xcolor}
\usepackage{fontawesome5}

\renewcommand{\arraystretch}{1.2} 
\newcommand{\cmark}{\textcolor{green!70!black}{\faCheckCircle}}
\newcommand{\tmark}{\textcolor{orange!80!black}{\faAdjust}}
\newcommand{\xmark}{\textcolor{red!70!black}{\faTimesCircle}}

 \usepackage[markup=underlined]{changes}

\usepackage{multirow}    
\usepackage{amssymb}     
\usepackage{booktabs}    

\usepackage{balance}

%% file: abstract.tex
Decompilers are fundamental tools for critical security tasks, from vulnerability discovery to malware analysis, yet their evaluation remains fragmented. Existing approaches primarily focus on syntactic correctness through synthetic micro-benchmarks or subjective human ratings, failing to address real-world requirements for semantic fidelity and analyst usability. We present \textbf{\sysname}, the first comprehensive framework that enables effective evaluation of decompilers in reverse engineering workflows through three key components: \textit{real-world function extraction} (comprising 23,400 functions from 130 real-world programs), \textit{runtime-aware validation}, and \textit{automated human-centric assessment} using LLM-as-Judge to quantify the effectiveness of decompilers in reverse engineering workflows.
Through a systematic comparison between six industrial-strength decompilers and six recent LLM-powered approaches, we demonstrate that LLM-based methods surpass commercial tools in code understandability despite 52.2\% lower functionality correctness.
These findings highlight the potential of LLM-based approaches to transform human-centric reverse engineering.
We open source \sysname\footnote{\url{https://github.com/Jennieett/DecompileBench}} to provide a framework to advance research on decompilers and assist security experts in making informed tool selections based on their specific requirements.

%% file: body.tex
\section{Introduction}
\label{sec:intro}

Modern software security fundamentally depends on understanding binary code. From identifying critical vulnerabilities in the network infrastructure to analyzing sophisticated malware, security analysts rely on decompilers to bridge the semantic gap between low-level machine instructions and human-comprehensible program logic. As a crucial line of defense against evolving cyber threats, these tools must not only faithfully recover program semantics, but also generate output that facilitates rapid analysis under real-world time constraints.

Two transformative forces have reshaped the decompilation landscape. First, modern software has grown increasingly complex and aggressive compiler optimizations systematically erase high-level semantics, making decompilation increasingly harder. Second, large language models have emerged as a disruptive force in decompilation. Reverse engineers now experiment with general-purpose models like GPT-4 for reverse engineering tasks~\cite{kwiatkowskiJusticeRageGepetto2025,reveng.aiRevEngAIReverseEngineering,binaryninjaBinaryNinjaSidekick}, while specialized models such as LLM4Decompile~\cite{tanLLM4DecompileDecompilingBinary2024} and MLM~\cite{ascendgraceMachineLanguageModel} focus on neural decompilation. These LLM-powered approaches show promise in generating more readable code by learning high-level programming patterns.

However, these advances bring new challenges. While LLMs-powered approaches produce visually coherent code, reverse engineers question whether these decompilations preserve the semantic behaviors crucial for security analysis. This uncertainty reveals a critical gap in how we evaluate decompilers for real-world security tasks.



\textbf{Crisis of Functionality Validation in Real-World Scenarios}.
Current decompiler evaluation approaches suffer from fundamental limitations. Most rely on artificially constructed programs~\cite{caoEvaluatingEffectivenessDecompilers2024} (e.g., Csmith-generated code~\cite{yangFindingUnderstandingBugs2011}) that lack the complexity of production software. 
When validating these decompilers, symbolic execution methods fail due to path explosion, while unit tests~\cite{tanLLM4DecompileDecompilingBinary2024,armengol-estapeExeBenchMLscaleDataset2022} only check input-output equivalence, thus ignoring critical behaviors like global state changes, heap manipulation, and exception handling that security analysts must trace when hunting vulnerabilities.

\textbf{Deficits of Automated Understandability Assessment}.
Current methods for evaluating decompiler output quality suffer from two key limitations. Traditional automated metrics like lines of code (LOC) and variable count~\cite{yangFindingUnderstandingBugs2011,caoEvaluatingEffectivenessDecompilers2024} miss crucial semantic properties such as meaningful variable names and logical clarity. Human studies~\cite{endersDewolfImprovingDecompilation2023,caoEvaluatingEffectivenessDecompilers2024,huDeGPTOptimizingDecompiler2024}, while providing valuable insights, lack sufficient scale across diverse compilation settings. This forces practitioners to choose between oversimplified metrics or expensive expert evaluations, neither suitable for assessing LLM-based decompilation approaches.


\begin{table*}[th!]
  \centering
  \renewcommand{\arraystretch}{1.2}
  \caption{Decompiler Evaluation Benchmark Matrix. 
  Symbol: \cmark \ (full support), \tmark \ (partial), \xmark \ (none).
  }
  \resizebox{0.98\textwidth}{!}{%
  \begin{tabular}{c c c c c c c c c}
    \toprule \multirow{2}{*}{\makecell{Dimensions}} & \multicolumn{8}{c}{Research Works} \\
    \cmidrule(r){2-9}                                       & \makecell{D-Helix \\ \cite{zouDHelixGenericDecompiler2024}}                            & \makecell{SAILR \\ \cite{basqueAhoySAILRThere}}    & \makecell{LLM4Decompile \\ \cite{tanLLM4DecompileDecompilingBinary2024}} & \makecell{ISSTA'20 \\ \cite{liuHowFarWe2020}} & \makecell{SEC'24 \\ \cite{dramkoTaxonomyDecompilerFidelity}} & \makecell{DecGPT \\ \cite{wongRefiningDecompiledCode2023}} & \makecell{Dewolf \\ \cite{endersDewolfImprovingDecompilation2023}} & \textbf{Ours} \\
    \midrule Real-world Binaries                            & \xmark                            & \tmark & \tmark        & \xmark   & \tmark & \tmark & \tmark & \cmark        \\
    \rowcolor{gray!10} Recompilation                        & \tmark                            & \xmark & \cmark        & \tmark   & \xmark & \cmark & \xmark & \cmark        \\
    Functionality Validation                                & \tmark                            & \tmark & \tmark        & \tmark   & \tmark & \tmark & \tmark & \cmark        \\
    \rowcolor{gray!10} Readability Metrics                  & \xmark                            & \tmark & \tmark        & \xmark   & \tmark & \xmark & \cmark & \cmark        \\
    Optimization Levels                                     & \xmark                            & \cmark & \cmark        & \xmark   & \cmark & \cmark & \xmark & \cmark        \\
    \rowcolor{gray!10} Full Automation                      & \cmark                            & \cmark & \tmark        & \cmark   & \xmark & \tmark & \tmark & \cmark        \\
    LLM-Based Decompiler                                    & \xmark                            & \xmark & \cmark        & \xmark   & \xmark & \xmark & \xmark & \cmark        \\
    \Xhline{1pt}                                              
  \end{tabular}%
  }%
  \label{tab:full-comparison} 
\end{table*}

To bridge these existing gaps, we present \sysname, the first comprehensive evaluation framework that holistically integrates multiple key evaluation dimensions, as summarized in Table~\ref{tab:full-comparison}. Our framework advances the field through the following innovations.

\textbf{Reconstructing Validation based on Runtime Consistency}. 
We resolve the synthetic evaluation crisis through a dual-pronged approach. By establishing a production-grade assessment pipeline via the OSS-Fuzz~\cite{OSSFuzz}, one of the largest continuous fuzzing frameworks for open-source software, we obtain 23,400 real-world functions from 130 actively maintained projects. We then validate decompilation correctness not through isolated unit tests or symbolic traces, but via runtime behavioral consistency during the fuzzing campaign. Our framework dynamically substitutes original functions with decompiled versions while instrumenting full-program execution paths and checking functionality correctness by evidencing through runtime behavioral across the entire binary.

\textbf{Redefining Understandability with Task-Driven Metrics}.
To overcome the limitations of existing readability metrics, we develop a 12-dimensional assessment framework grounded in reverse engineering objectives. We employ LLM-as-a-Judge to perform scalable comparisons of decompiler outputs through security-task lenses, such as \textit{Control Flow Clarity} and \textit{Memory Layout Accuracy}.
This automated approach, validated against expert ratings with $\kappa$=0.778 agreement, not only ranks decompilers but predicts their effectiveness for specific aspects in aiding the real-world reverse engineering tasks for reverse experts.

In summary, our main contributions are:

\begin{itemize}
  \item \textbf{Real-World Evaluation Framework}. We develop a decompiler assessment framework to assess the decompiler from three aspects. By using real-world binary generation and runtime validation to overcome synthetic limitations, we provide reports on compiler and runtime aspects. We also use LLM-as-a-Judge for scalable decompilation assessment, replacing subjective evaluations and simplistic metrics, to conduct code quality-aspect evaluation and measure the helpfulness in human-centric reverse engineering.
  \item \textbf{Empirical Security Insights}. Our evaluation of 12 decompilers (6 traditional, 2 decompilation-specialized models, and 4 general-purpose LLMs) yields transformative insights. Our key findings contain Hex-Rays' 36.11\% functionality correctness drop under \texttt{-O3} optimizations, LLM's significant readability improvement over Hexrays despite lower recompilation rates. These findings reshape our understanding of decompiler capabilities in security-critical contexts.
  \item \textbf{Open-Source Benchmark}. We release \sysname to enable systematic decompiler evaluations for future research and help analysts choose context-appropriate tools.
\end{itemize}

\section{Related Works}


\subsection{Decompiler Evaluation}

Decompilers are crucial in reverse engineering, converting binary executables into human-readable high-level source code. This task is challenging because compilers remove key information during compilation, such as variable types and control structures. Decompilers use predefined rules and heuristics to reconstruct these details~\cite{basqueAhoySAILRThere,shoshitaishviliStateArtWar}, leading to significant performance variability based on their rule implementations. This variability highlights the necessity for systematic evaluation frameworks to assess decompiler quality,  particularly in terms of functionality correctness and readability.

\subsubsection{Functionality Correctness Evaluation}

Functionality correctness evaluation aims to verify whether the decompiled code matches the original program in functionality. Prominent methods include Equivalence Modulo Inputs (EMI) testing~\cite{liuHowFarWe2020}, which compares global variable checksums between original and decompiled code executions. However, EMI relies on Csmith-generated~\cite{yangFindingUnderstandingBugs2011} test cases, which may oversimplify real-world complexity. To address this, Dsmith~\cite{caoEvaluatingEffectivenessDecompilers2024} preserves intricate control and data flows by focusing on runtime-dependent variables, while D-Helix\cite{zouDHelixGenericDecompiler2024} leverages real-world binaries from GitHub. 
Symbolic execution tools like Diff~\cite{kimTestingIntermediateRepresentations2017}, Alive~\cite{lopesProvablyCorrectPeephole2015}, Alive2~\cite{lopesAlive2BoundedTranslation2021}, and SYMDIFF~\cite{zouDHelixGenericDecompiler2024} further enhance correctness verification by analyzing intermediate representations (IR) and mapping symbolic models. Despite its advantages, symbolic execution faces challenges such as path explosion, which can compromise precision.

\subsubsection{Readability Assessment}

Readability assessment evaluates the clarity and understandability of decompiled code, as the goal is to produce high-level representations for human understandability. Metrics such as Cyclomatic Complexity, Lines of Code, number of goto statements, and variable counts are commonly used to assess control and data flow characteristics~\cite{caoEvaluatingEffectivenessDecompilers2024}. Techniques like DREAM's pattern-independent algorithm~\cite{basqueAhoySAILRThere} and RevNG-C's "Control Flow Combing" aim to reduce control complexity and eliminate goto statements~\cite{gussoniCombDecompiledCode2020}. Human evaluation is also crucial. Many studies involve experienced analysts and online platforms comparing decompiler outputs based on control flow structure and code patterns~\cite{yakdanHelpingJohnnyAnalyze2016, endersDewolfImprovingDecompilation2023, caoEvaluatingEffectivenessDecompilers2024, dramkoTaxonomyDecompilerFidelity,eomR2IRelativeReadability2024}. 

\subsection{Machine Learning for Decompilation}

Machine learning has progressively advanced decompilation, starting with basic RNNs for binary-to-C translation~\cite{katzUsingRecurrentNeural2018} and evolving with NMT techniques that recover semantic details like variable names and types~\cite{katzNeuralDecompilation2019,liangSemanticsRecoveringDecompilationNeural2021,hosseiniRetargetableDecompilationUsing2022}. Recent LLM-based frameworks like DecGPT~\cite{wongRefiningDecompiledCode2023} and DeGPT~\cite{linMoDeGPTModularDecomposition2024} employ hybrid methods, including static and dynamic phases, to optimize decompilation. LLM4Decompile~\cite{tanLLM4DecompileDecompilingBinary2024} fine-tunes LLMs to align decompiled and original code, and ReSym~\cite{xieReSymHarnessingLLMs2024} recovers variable semantics using LLMs combined with reasoning systems. Advanced foundation models like GPT~\cite{openaiGPT4TechnicalReport} and DeepSeek~\cite{deepseek-aiDeepSeekV3TechnicalReport2024} have demonstrated an enhanced understanding of decompiled code.

\subsection{LLM as a Judge}
The recent advancements in large language models (LLMs) have led to the introduction of the `LLM-as-a-judge' concept~\cite{zhengJudgingLLMasajudgeMTBench2023}, which utilizes the capabilities of LLMs to score and rank multiple candidates~\cite{liGenerationJudgmentOpportunities2025}. 
LLMs are capable of evaluating various aspects such as reliability~\cite{chengEvaluatingHallucinationsChinese2023}, helpfulness~\cite{leeRLAIFScalingReinforcement2023}, relevance~\cite{luDeepCRCEvalRevisitingEvaluation2025}, and conducting multi-aspect assessments across diverse applications~\cite{yuKIEvalKnowledgegroundedInteractive2024}. 
To perform pair-wise comparison and provide comprehensive feedback~\cite{shenComparativeAnalysisListwise2025} without positional bias, techniques such as CoT-like prompting~\cite{zhuStarling7BImprovingHelpfulness2024} and output-swapping~\cite{zhengJudgingLLMasajudgeMTBench2023} are proposed, with tournament-based methods~\cite{leeAligningLargeLanguage2024} further accelerating evaluations.


\section{Methodology}


To evaluate decompilation in real-world settings, we use the OSS-Fuzz to construct dataset.
We further develop an evaluation framework that comprehensively evaluates the decompiler's performance from three aspects cared for by the end users.

\subsection{Dataset Construction}
\label{subsec:dataset-engine}

\begin{figure*}[t!]
\centering
    \includegraphics[width=0.8\linewidth]{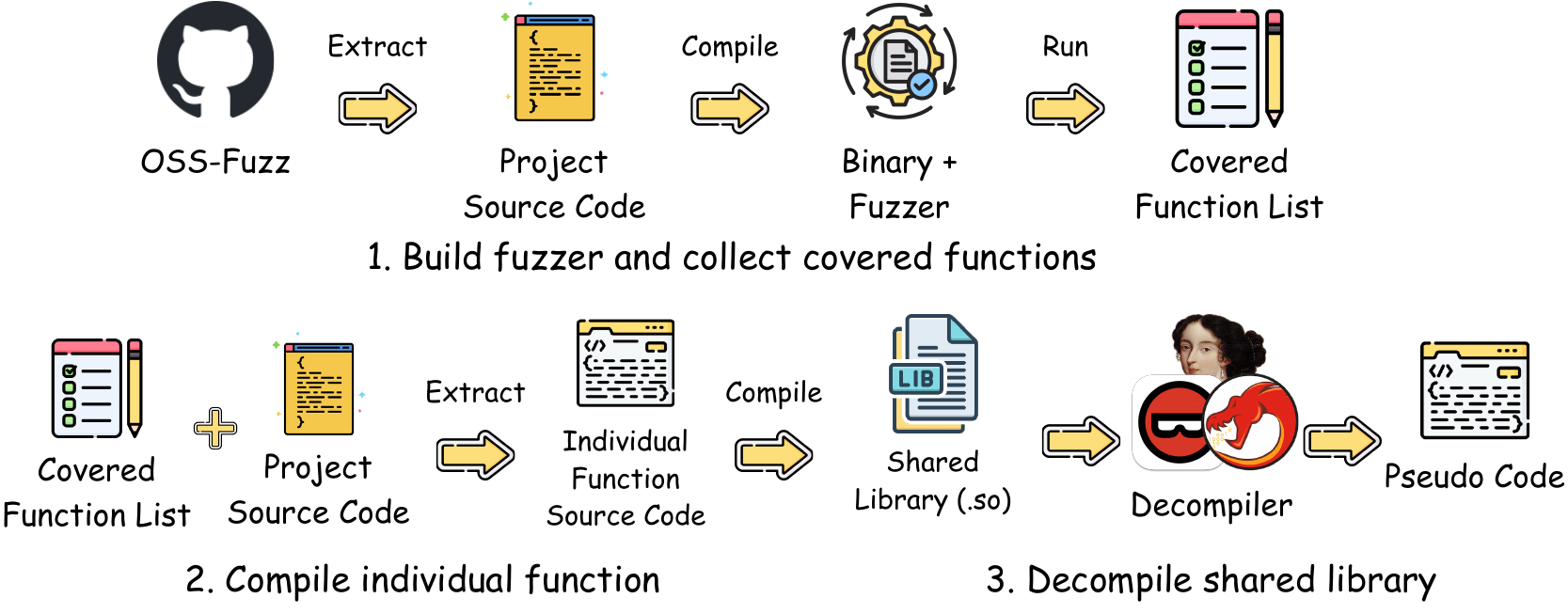}
    \caption{Overview of dataset construction process.}
\label{fig:dataset_engine}
\end{figure*}


\begin{figure*}[ht!]
  \centering
      \includegraphics[width=0.98\linewidth]{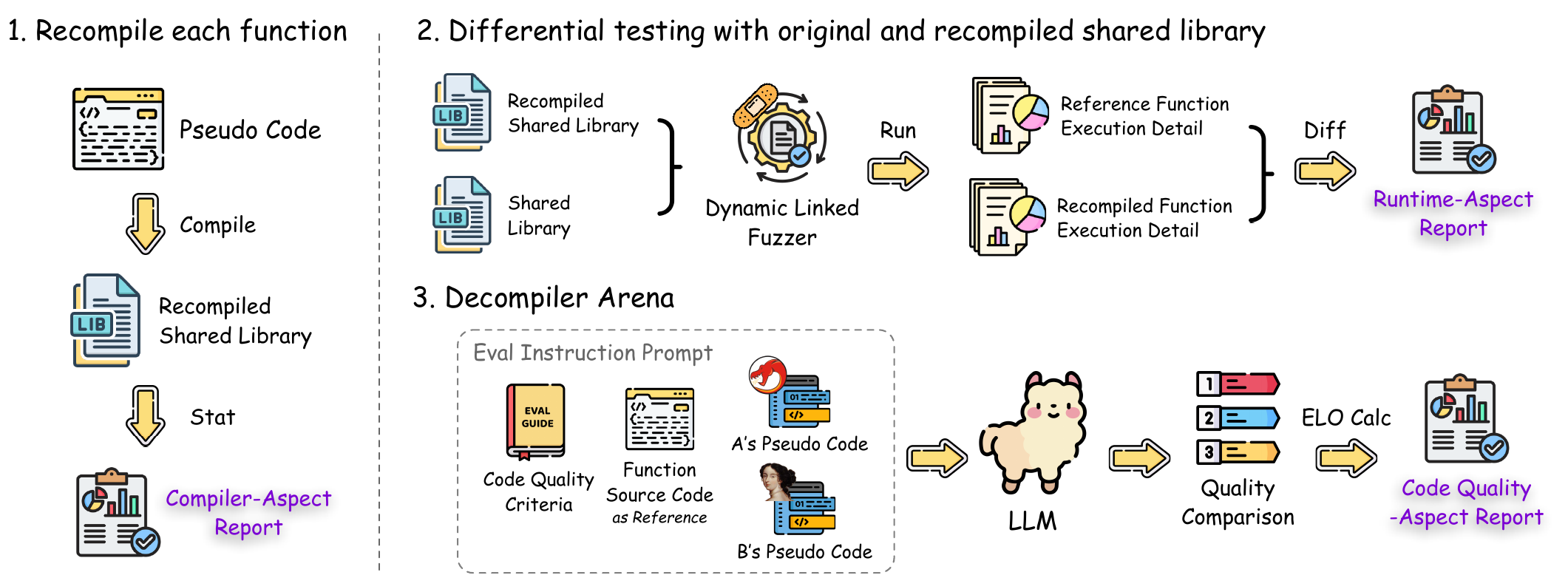}
      \caption{Three-dimensional evaluation framework for decompiler assessment: Successful recompilation rate, Runtime behavior consistency, and LLM-assessed code quality.}
  \label{fig:eval_metrics}
\end{figure*}

The dataset construction stage, shown in Figure~\ref{fig:dataset_engine}, begins with the extraction of source code from OSS-Fuzz projects~\cite{OSSFuzz}. OSS-Fuzz is a platform that provides continuous fuzzing for well-known open-source software. These projects are configured and compiled with Clang's coverage sanitizer enabled, producing executable fuzzers and initial seeds as fuzzing input. By running these fuzzers fed with seeds as input, we utilize Clang's coverage sanitizer to identify functions covered during execution. For each covered function, we use \texttt{clang-extract}~\cite{SUSEClangextract2024} to extract its implementation along with all dependencies, such as called function signatures, and used macro definitions. This allows us to compile individual functions into standalone binaries (shared object files, \texttt{.so}). Finally, we decompile the desired function from these binaries using multiple decompilers to obtain the decompiled code for further evaluation. We leave the detailed compile options and decompiler configurations for Section~\ref{sec:compilation-decompilation-detail}.

\subsection{Evaluation Aspect}

Following the decompilation process, we evaluate the outputs across three aspects, as shown in Figure~\ref{fig:eval_metrics}:  the compiler aspect, the runtime aspect, and the code quality aspect. These aspects measure the re-compilation rate, functionality consistency, and the readability and practical utility of the decompiled code for reverse engineers.

\subsubsection{Compiler-Aspect Report}
The usability of decompiled code hinges first and foremost on its ability to meet compiler requirements, including correct language syntax and typing. 
To achieve this, we combine the decompiled code with the previously extracted include directives and attempt to recompile it. The result of this process is captured in a \textit{Compiler-Aspect Report}, which quantifies the recompilation success rate. A high success rate indicates that the decompiler has preserved the essential syntactic structure and dependencies of the original code, making the decompiled output both functionally viable and practically useful.

\subsubsection{Runtime-Aspect Report}
\label{subsubsec:runtime-aspect-report}

While successful recompilation ensures syntactic validity, it does not guarantee that the decompiled code maintains the original program's functionality. Previous methods, such as symbolic execution~\cite{caoEvaluatingEffectivenessDecompilers2024} and unit testing~\cite{tanLLM4DecompileDecompilingBinary2024}, have attempted to verify decompilation accuracy but face inherent limitations. Symbolic execution struggles with path explosion in real-world programs, and unit testing focuses mainly on output equivalence, often missing complex inter-function dependencies involving global variables and multi-level pointers.

Our approach introduces an innovative side-effect consistency paradigm inspired by duck-typing~\cite{pythondocGlossary} in dynamic languages. We propose that two implementations are functionally equivalent if they produce identical side effects on the program's execution environment. Practically, a decompiled function is likely correct if substituting it into the original program results in identical execution characteristics across all components.
The implementation details of non-interfering function substitution are detailed in Section~\ref{sec:function-substitution}.

In the evaluation, we operationalize this principle through branch coverage consistency analysis using Clang's SanitizerCoverage~\cite{llvmSanitizerCoverageClangDocumentation} and leverage OSS-Fuzz's infrastructure in three critical phases to conduct the verification: 

\begin{enumerate}
  \item Reference Function Profiling: We substitute the target function with its original source-compiled version (a shared library), and we execute the modified program with the seed corpus collected in Section~\ref{subsec:dataset-engine} as input, but without entering fuzzing iterations. This process collects full-program branch coverage through sanitizer instrumentation. This establishes the reference function execution profile containing ground truth coverage metrics.
  \item Decompiled Code Profiling: Following the same substitution paradigm, we substitute the target function with its decompiled version while retaining the same seed corpus, then perform identical execution to generate the recompiled function execution profile.
  \item Differential Analysis: After obtaining the two profiles, we compare the reference and recompiled function execution profile to check whether the decompilation preserves the control flow patterns during the whole binary execution. Here, preserving control flow patterns requires strict equivalence in execution counts of conditional statements (\texttt{if/for/while}) and boolean outcomes distribution for conditional branches (\texttt{true/false} ratios).
\end{enumerate}

Moreover, this validation method can be seamlessly extended by plugging in existing instrumentation framework for finer granularity---additional comparison on stdout/stderr output, local variables~\cite{fioraldiUseLikelyInvariants2021} and operands in comparison expression~\cite{aschermannREDQUEENFuzzingInputtoState2019}.

\subsubsection{Code Quality-Aspect Report}
To systematically assess decompiled code quality, we develop a dual-faceted evaluation framework focusing on \textit{readability} (syntactic comprehension) and \textit{helpfulness} (semantic reconstruction).
We extend prior works~\cite{dramkoTaxonomyDecompilerFidelity,caoEvaluatingEffectivenessDecompilers2024} and establish 12 fine-grained evaluation criteria spanning five readability aspects (e.g., type system consistency), five helpfulness dimensions (e.g., identifier semantics), and two hybrid criteria affecting both characteristics detailed in Table~\ref{tab:code-quality-aspects}.

\begin{table*}[!t]
\centering
\scalebox{0.68}{
\begin{tabular}{c|l|l|l}
  \Xhline{1pt}
  \textbf{Category} & \textbf{Subclass} & \textbf{Explanation} & \textbf{Example} \\ 
  \Xhline{1pt}
  \multirow{6}{*}{Readability} & {Typecast Correctness} & Redundant/incorrect type casts obscure intent & \texttt{exit((long long)"Invalid size")} \\
  \cline{2-4}
  & {Literal Representation} & Non-idiomatic literals hinder understanding & numeric 2685 instead of string literal ``\texttt{\textbackslash{}n}'' \\
  \cline{2-4}
  & {Control Flow Clarity} & Complex pointer dereferencing & \makecell[l]{%
    \texttt{%
      for
      \textcolor{cyan}{(}%
        i = a2; a1 != %
        \textcolor{teal}{(}%
          *((\_QWORD *)(i+64))%
        \textcolor{teal}{)}%
        ;%
    } \\ 
    \texttt{%
        \: i = *
        \textcolor{red}{(}%
          (\_QWORD *)(i+64)
        \textcolor{red}{)}%
      \textcolor{cyan}{)}%
    }
  } \\
  \cline{2-4}
  & {Decompiler Macros} & Non-standard macros violate conventions & \texttt{LOWWORD(v5)} \\
  \cline{2-4}
  & {Return Behavior} & Altered return expressions change logic & \texttt{return \_\_readfsqword(0x28u)\^{}v3} \\
  \hline
  \multirow{5}{*}{Helpfulness} & {Identifier Meaning} & Generic names reduce semantic value & Using \texttt{v4} instead of \texttt{buffer} \\
  \cline{2-4}
  & {Identifier Accuracy} & Misleading variable semantics & \texttt{error\_flag} vs \texttt{total\_count} \\
  \cline{2-4}
  & {Symbolic Values} & Hardcoded values reduce clarity & 8 instead of \texttt{sizeof(long)} \\
  \cline{2-4}
  & {Function Correctness} & Core functionality recovery failure & Overly too complex logic to comprehend \\
  \cline{2-4}
  & {Function Precision} & Approximate functionality recovery & MD5 identified as SHA-256 implementation \\
  \hline
  \multirow{3}{*}{\makecell[c]{
    Readability \\
    {\small \&} \\
    Helpfulness
  }} & {Dereference Readability} & Opaque pointer arithmetic & \texttt{((\_QWORD *)v5 + 8)} \\
  \cline{2-4}
  & {Memory Layout} & Failed type inference &  
  \makecell[l]{%
    \texttt{%
      \textcolor{cyan}{(}%
        *%
        \textcolor{teal}{(}%
          void (\_\_stdcall **)%
          \textcolor{blue}{(}%
            DWORD%
          \textcolor{blue}{)}%
        \textcolor{teal}{)}
    } \\ 
    \texttt{%
        \quad%
        \textcolor{red}{(}%
          *%
          (\_DWORD *)lpD3DDev\_1+68
        \textcolor{red}{)}%
      \textcolor{cyan}{)}%
      (pD3DDev\_1);
    }
  }
  \\
  \Xhline{1pt}
\end{tabular}
}
\caption{Code Quality Aspects: 12 aspects categorized by readability, helpfulness, and both.}
\label{tab:code-quality-aspects}
\end{table*}


The evaluation process employs Qwen-2.5-Coder-32B~\cite{qwenQwen25TechnicalReport2024} to conduct aspect-granular comparisons: For each pair of decompilers (A vs. B), the LLM \romannumeral1) uses the reference function code as the ground truth, \romannumeral2) assesses both outputs against every criterion listed in Table \ref{tab:code-quality-aspects}, and \romannumeral3) selects a winner for each criterion, accompanied by a detailed justification, output in a predefined JSON format. These pairwise outcomes are then used to dynamically compute Elo scores, which serve as a quantitative representation of each decompiler's code quality.
To balance thoroughness and efficiency in our evaluation, we introduce a probability-based sampling strategy (detailed in Section~\ref{sec:llm-as-a-judge-sampling-strategy}) that prioritizes comparisons between decompilers with similar Elo ratings. This targeted approach increases the density of comparisons among closely ranked models, enabling more precise discrimination of subtle performance differences while maintaining broad evaluation coverage.

\section{Evaluation}

Our evaluation encompasses six traditional decompilers, representing both commercial and open-source solutions for reverse engineering, the description and decompilation technical details are presented in Section~\ref{sec:decompiler-details}.
For the emerging paradigm of LLM-based decompilation, we evaluate both domain-specific and general-purpose models: LLM4Decompile is tested using its official prompt template, and MLM is assessed through its public API service. The general models (GPT-4o-mini, GPT-4o, Qwen2.5-Coder-32B-Instruct, Claude-3.5-Sonnet, and DeepSeek-V3) are instructed to refine Hex-Rays outputs using task-specific guidelines and three illustrative examples as few shots. All LLM evaluations employ generation parameters with a maximum token limit of 8,192, temperature of 0.7, and $\mathit{top}_p$ of 1.0.
We use Clang 18 on Ubuntu 22.04 to compile. We build our compilation service upon official OSS-Fuzz with modifications to achieve function substitution and evaluation metrics collection. 



To evaluate whether LLMs can effectively capture human judgments on decompiled code quality, we employed Cohen's kappa ($\kappa$) to measure the agreement between LLM and human ratings. We describe the detail in Section~\ref{sec:cohen-kappa-detail}.

\subsection{Compiler- and Runtime-Aspect Analysis}

\begin{table*}[h!]
    \centering
\resizebox{0.85\textwidth}{!}{
    \begin{tabular}{l|c|c|c|c|c|c|c|c|c|c|c|c}
        \Xhline{1pt}
        \multirow{3}{*}{\textbf{Decompiler}} 
        & \multicolumn{6}{c|}{\textbf{Recompile Success Rate}} 
        & \multicolumn{6}{c}{\textbf{Coverage Equivalence Rate}} \\
        \cline{2-13}
        & \textbf{O0} & \textbf{O1} & \textbf{O2} & \textbf{O3} & \textbf{Os} & \textbf{Avg} 
        & \textbf{O0} & \textbf{O1} & \textbf{O2} & \textbf{O3} & \textbf{Os} & \textbf{Avg} \\
        \Xhline{0.9pt}
        Angr      & 0.309 & 0.232 & 0.190 & 0.181 & 0.191 & 0.221 
                  & 0.187 & 0.153 & 0.124 & 0.116 & 0.118 & 0.140 \\
        Binja     & 0.274 & 0.246 & 0.229 & 0.215 & 0.224 & 0.238 
                  & 0.167 & 0.153 & 0.137 & 0.129 & 0.138 & 0.145 \\
        Dewolf    & 0.225 & 0.203 & 0.214 & 0.204 & 0.222 & 0.213 
                  & 0.125 & 0.120 & 0.113 & 0.111 & 0.118 & 0.117 \\
        Ghidra    & 0.524 & 0.421 & 0.395 & 0.377 & 0.353 & 0.413 
                  & 0.374 & 0.294 & 0.256 & 0.241 & 0.228 & 0.278 \\
        Hex-Rays   & 0.706 & 0.573 & 0.558 & 0.513 & 0.565 & \textbf{0.583} 
                  & 0.523 & 0.430 & 0.392 & 0.361 & 0.400 & \textbf{0.418} \\
        Retdec    & 0.402 & 0.349 & 0.337 & 0.329 & 0.355 & 0.354 
                  & 0.185 & 0.160 & 0.143 & 0.137 & 0.149 & 0.155 \\
        MLM       & 0.335 & 0.321 & 0.313 & 0.311 & 0.314 & 0.319 
                  & 0.216 & 0.205 & 0.188 & 0.191 & 0.198 & 0.200 \\
        LLM4Decompile & 0.285 & 0.270 & 0.257 & 0.250 & 0.256 & 0.264 
                      & 0.192 & 0.177 & 0.153 & 0.150 & 0.147 & 0.164 \\
        Qwen2.5-Coder-32B & 0.659 & 0.528 & 0.515 & 0.480 & 0.526 & 0.542 
                          & 0.385 & 0.302 & 0.264 & 0.249 & 0.281 & 0.296 \\
        Deepseek-V3 & 0.663 & 0.539 & 0.531 & 0.492 & 0.539 & 0.553 
                    & 0.403 & 0.328 & 0.316 & 0.283 & 0.313 & 0.329 \\
        GPT-4o-mini & 0.658 & 0.591 & 0.559 & 0.531 & 0.572 & \underline{0.582} 
                    & 0.296 & 0.269 & 0.231 & 0.210 & 0.244 & 0.254 \\
        GPT-4o & 0.649 & 0.553 & 0.537 & 0.509 & 0.548 & 0.559 
               & 0.410 & 0.352 & 0.323 & 0.312 & 0.334 & \underline{0.346} \\
        Claude-3.5-Sonnet$^*$ & 0.413 & 0.339 & 0.308 & 0.329 & 0.360 & 0.350 
               & 0.277 & 0.224 & 0.196 & 0.196 & 0.239 & 0.227 \\

        \Xhline{1pt}
    \end{tabular}
    }
    \caption{Recompile success rate and coverage equivalence rate of various decompilers across different compiler optimization levels. $^*$Tested on a randomly sampled dataset comprising 1/5 dataset due to the high cost. }
    \label{tab:Combined_Results}
\end{table*}

Our evaluation across the compiler aspect and runtime aspect are revealed through two metrics, \textit{Recompile Success Rate} (RSR) and \textit{Coverage Equivalence Rate} (CER), as the result shown in Table~\ref{tab:Combined_Results}. 

Hex-Rays is recognized as the leading traditional decompiler, achieving the highest single optimization-level RSR of 0.706 at \texttt{-O0}, and excelling in averaged metrics with an RSR of 0.583 and a CER of 0.417. 
Ghidra ranks second among traditional tools, 
establishing itself as the best open-source decompiler. In contrast, dewolf performs the worst in both metrics, as it is designed to prioritize readability for users.

Among LLM-based decompilation approaches, GPT-4o shows the highest semantic fidelity with a CER of 0.346, while GPT-4o-mini closely matches Hex-Rays' syntactic recovery capability with an RSR of 0.582. However, these LLM-enhanced results still fall short of Hex-Rays' original metrics, with RSR lower by 0.2-45.3\% and CER lower by 17.2-52.2\%. This is because the LLMs prioritize readability over strict compiler compatibility.
General-purpose models outperform decompilation-specialized models by 69.9-120.8\% in the RSR metric and by 27.3-111.6\% in the CER metric. This performance gap may be due to the \textit{temporal advantage} in their development timelines. The swift advancement of general large language models, which were released 9-12 months after the decompilation-specialized models, indicates that they can exceed the capabilities of specialized models within short technological windows.

Our experiments also reveal some key insights. First, CER generally correlates with RSR, as runtime validation inherently depends on successful recompilation. However, exceptions arise-RetDec achieves a higher RSR (0.354) than MLM (0.319) but a lower correct execution rate (CER) (0.155 vs. 0.200), indicating RetDec's prioritization of compilation over functionality. GPT-4o-mini achieves superior RSR in \texttt{-O1} to \texttt{-Os} optimizations by employing ad-hoc adaptations, such as renaming Hex-Rays' incomplete function calls from \texttt{xmlSAX2ErrMemory()} to \texttt{handle\_xml\_memory\_error()}. This undefined function triggers a warning rather than an error during compilation into a shared library. Though this ensures compilation success, it results in runtime failures and inferior CER compared to Hex-Rays.

Second, both RSR and CER decline consistently from \texttt{-O0} to \texttt{-O3}, with \texttt{-Os} outperforming \texttt{-O3} and closely aligning with \texttt{-O2}. This trend highlights the increasing difficulty of recovering compiler-compatible code under aggressive optimization.
Notably, Hex-Rays' RSR declines by 27.3\% across this spectrum (from 0.706 to 0.513) and fails to achieve a CER above 50\% at \texttt{-Os}. This highlights the persistent challenges of semantic recovery and emphasizes the need for significant advancements in both LLM-augmented and traditional decompilation approaches.

\subsection{Code Quality Analysis}
Our automated code quality assessment highlights some key insights. First, LLM-generated decompiled code consistently surpasses traditional decompilers in quality. This is true for specialized models, which benefit from training focused on readability, and general-purpose LLMs, which improve through few-shot guidance on structured code corpora. 
Second, \textsc{MLM} excels in enhancing readability, achieving an ELo score of 1581 compared to Hex-Rays' 1162, particularly excelling in \textit{Control Flow Clarity} and \textit{Literal Representation Correctness} (see Section~\ref{sec:control_flow_case}). However, LLMs sometimes experience hallucinations during variable inference, leading to minor semantic inaccuracies that limit improvements in \textit{Identifier Name Correctness} and \textit{Typecast Correctness}. 
Third, among traditional decompilers, Hex-Rays scores higher in readability but faces Macro Conformity issues due to the usage of nonstandard C macros, while RetDec underperforms across all metrics. 
Additionally, the ELo rankings for individual aspects closely match the overall rankings, indicating that decompilers rarely excel in some areas while performing poorly in others. This consistency underscores the reliability of ELo as a comprehensive measure of decompilation quality.

To validate the ecological validity of our LLM-based assessments, we conducted human evaluations through dual expert annotation. Two independent evaluators assessed 30 randomly selected samples, achieving an overall Cohen's kappa coefficient of 0.778. Detailed agreement statistics across quality dimensions appear in Section~\ref{sec:agreement_analysis}.
\begin{figure}[t!]
  \centering
  \includegraphics[width=1.0\linewidth]{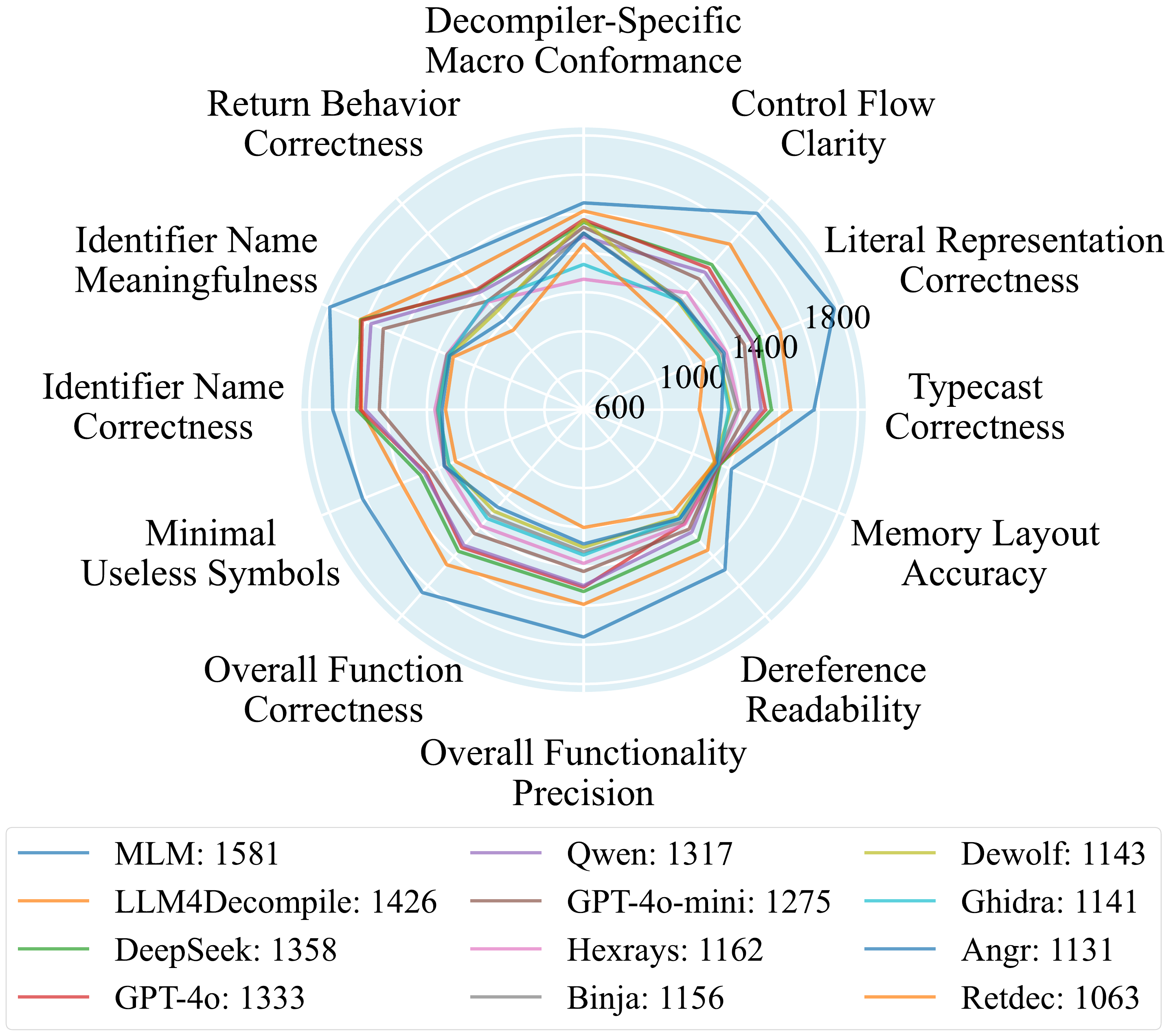}
  \caption{
    Comparison of code quality across twelve dimensions using Elo scores. 
    The average Elo score across all dimensions is shown in the bottom legend.
    The scores are relative within each dimension, with higher scores indicating a higher win rate. 
    Note that absolute scores across different dimensions are not directly comparable.}
  \label{fig:arena_radar
  }
\end{figure}

\subsection{From Heuristic to Neural}
We evaluate traditional and LLM-based decompilers through recompilation errors, comparing their strengths and limitations to guide decompiler selection for different use cases.

\subsubsection{Semantic Fidelity Tradeoffs}  
Our empirical study exposes fundamental divergences in semantic preservation between rule-based and neural decompilation paradigms.
To systematically quantify causes of decompilation errors, we extend the error types from previous work~\cite{caoEvaluatingEffectivenessDecompilers2024} with new types identified in our compilation pipeline, culminating in 15 error types whose empirical distributions across different compilers are statistically profiled in Section~\ref{subsec:error-analysis}.

Traditional tools demonstrate strict adherence to low-level accuracy through deterministic pointer arithmetic (\texttt{*((\_DWORD *)\&ses + 4)}), yet manifest systematic limitations like
type safety violations via unsafe casts (\texttt{*(\_\_m128i *)(a2 + 8)})
or 
const qualification breaches during pointer dereferencing, e.g. perform \texttt{*s = 0} which is declared as readonly \texttt{const char* s}.

In contrast, LLM-based approaches (GPT-4o/MLM) achieve higher AST readability at the cost of introducing novel failure modes: 
Hallucinated type constructs (e.g., synthetic undefined \texttt{archive\_t} replacing \texttt{\_\_int64}); 
Speculative header injections (\texttt{\#include "sudo\_debug.h"} which is not in the context); 
Critical parameter omission in function ABIs, breaking function invocations by removing seemingly unused parameters (e.g., \texttt{float a3}), even when such parameters are critical to runtime compatibility. 

\begin{figure}[t!]
  \centering
  \includegraphics[width=1.0\linewidth]{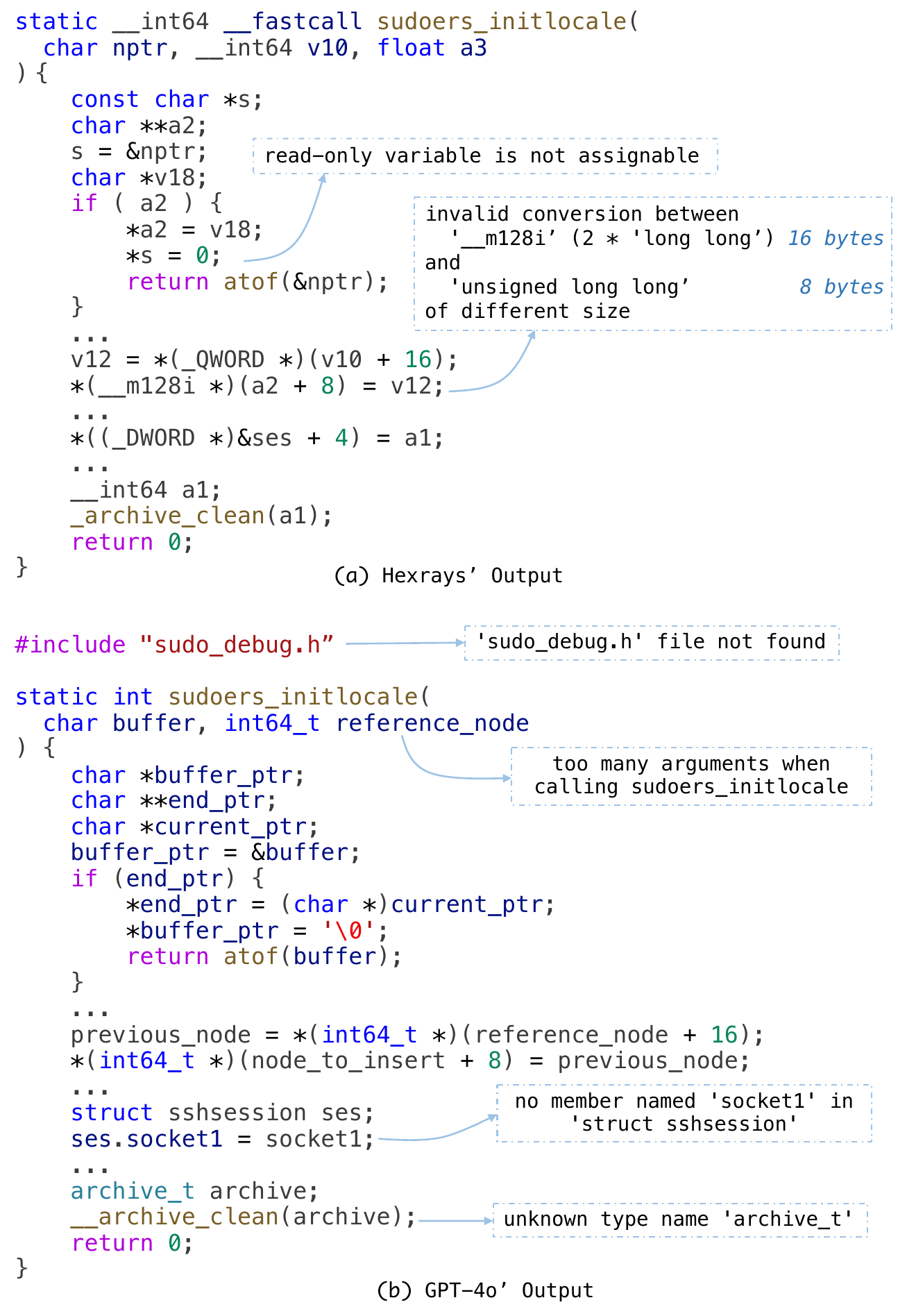}
  \caption{Output of Hex-Rays and GPT-4o. The error reports highlight where one decompiler fails and the other succeeds, demonstrating their pros and cons.}
  \label{fig:error_compare}
\end{figure}

The neural paradigm particularly struggles with pointer arithmetic resolution, as evidenced in Figure~\ref{fig:error_compare} where GPT-4o generates invalid struct member access (\texttt{ses.socket1}) versus Hex-Rays' bit-precise implementation. Hybrid approaches show promise in bridging this gap, as demonstrated by LLM-corrected type casts (\texttt{*(int64\_t *)(node\_to\_insert + 8)}) and relaxed type constraints (e.g., redefining \texttt{*buffer\_ptr} as mutable \texttt{char}), surpassing traditional tools' output validity.


\subsubsection{Scenario-Driven Tool Choosen}

These findings reveal a trade-off between readability and correctness, suggesting the need to combine LLMs' contextual flexibility with traditional tools' rigorous type-checking in practical applications.

For reliability-critical scenarios like performance analysis and debugging, established decompilers such as Hex-Rays and Ghidra remain preferable due to their semantically accurate and dependable outputs, even if they are sometimes less user-friendly. Conversely, in reverse engineering where quick comprehension is crucial, such as malware detection, LLM-based decompilers are preferred. 
Especially those models fine-tuned for clarity, like MLM, offer enhanced readability that significantly aids analysis. 
Ultimately, the choice of decompilation technique should be guided by the specific analytical objectives, balancing stringent accuracy with human interpretability.

\section{Conclusion}
We present \sysname, a comprehensive decompiler evaluation framework that addresses real-world assessment challenges through three key innovations: production-grade datasets from OSS-Fuzz, runtime behavior validation, and LLM-powered code quality analysis. Our evaluation of 12 decompilers highlights critical trade-offs: traditional tools such as Hex-Rays achieve a 58.3\% average recompilation success rate, whereas LLM-based approaches like MLM excel in code quality, offering superior control flow clarity and more meaningful identifier naming.

This work establishes new standards for decompiler evaluation in security-critical scenarios and highlights the need for hybrid approaches combining neural-based decompilers with rule-based approaches. Our framework enables informed tool selection for reverse engineers while guiding future research toward balancing reliability with human-centric readability for reverse engineering.


\section{Limitation}




\subsection{Error Localization}
Our analysis evaluates recompilation failures as semantic fidelity indicators for decompilation quality. While runtime differential coverage metrics detect functional discrepancies between original and decompiled outputs, opaque dependencies in shared libraries hinder precise error localization.

\subsection{Root-Cause Diagnostic}
Our methodology focuses on \textit{where} decompilers underperform, but the complexity of decompilation obscures the \textit{why} behind these failures, preventing root-cause analysis. 
This underscores the need for further exploration, such as differential testing of heuristics, to improve the internal mechanisms of open-source decompilers, as demonstrated by studies like \cite{zouDHelixGenericDecompiler2024}.

\subsection{LLM-Augmented Reprocessing}
To address compilation errors with LLMs, we did not integrate Clang-extracted dependencies (e.g., typedefs, structs, macros, includes) into LLM inputs due to resource constraints. This integration could enable joint inference of type definitions and resolution of external calls. Moreover, feeding error information to LLMs and iteratively recompiling their outputs could enhance decompilers. However, due to time constraints, this remains a task for future work.

\subsection{Architecture-Specific Evaluation}
The evaluation is restricted to x86-64 architectures, despite platform-dependent variations in shared library behavior and compiler optimizations. For example, decompiled functions relying on x86-64-specific features like register usage or memory alignment may fail on ARM architectures.

\subsection{Limited Human Validation}
While achieving expert-LLM agreement ($\kappa$=0.778) on code quality metrics, the validation is limited to a small set of cases (n=30) evaluated by only two annotators. This narrow scope risks overlooking diverse human perspectives and potential `reward hacking', where models optimize for evaluation patterns rather than genuine quality improvements. Expanding to diverse architectures and a larger annotator pool would enhance the generalizability and robustness of our findings.


%% file: appendix.tex
\section{Decompiler Details}
\label{sec:decompiler-details}

This appendix provides description and technical details about each decompiler in our paper:

\begin{itemize}
   \item \textbf{Angr}: 
   Binary analysis framework utilizing proprietary AIL (angr Intermediate Language) combining symbolic execution with Value-Set Analysis (VSA).
   Executes decompilation through \texttt{angr.Project.Decompiler} class via direct invocation on target functions. Supports x86, ARM, and MIPS architectures.

   \item \textbf{Binary Ninja}: 
   Multi-stage decompilation platform with LLIL/MLIL/HLIL intermediate representations. Creates decompilation views via \path{LinearViewObject.single_function_language_representation(func, settings)}, constructing final output through iterative cursor object traversal.
   
   \item \textbf{Dewolf}: 
   Binary Ninja-based decompiler optimized for human-readable C code. Initializes decompilation process using \texttt{DecompilerPipeline.from\_strings()} that ingests control flow graph (CFG) and abstract syntax tree (AST) inputs, subsequently executing the pipeline for code generation.
   
   \item \textbf{Ghidra}: 
   Enterprise-grade system using P-Code intermediate language. Invokes decompilation through \texttt{FlatDecompilerAPI(flat\_api)} interface followed by explicit \texttt{decomp\_api.decompile(function)} calls in the API execution chain.
   
   \item \textbf{Hex-Rays}: 
   Industry-standard decompiler implementing \texttt{ida\_hexrays.decompile(func)} for single-step decompilation within IDA Pro's interactive environment. Features production-grade quality through mature pattern recognition.
   
   \item \textbf{RetDec}: 
   LLVM-based decompilation framework using multi-pass optimization. Executes standalone \texttt{retdec-decompiler} executable (located in installation directory) directly on target binaries to initiate the decompilation process.
\end{itemize}

\section{Compilation and Decompilation Detail}
\label{sec:compilation-decompilation-detail}

When compiling fuzzer and executable, we append \texttt{-Wl,--export-dynamic} to compile commands, enabling shared library to resolve symbols in the executable in function substitution~\ref{subsubsec:runtime-aspect-report}. 
When compiling target functions into shared libraries, we add \texttt{-fno-inline} to prevent function inlining that would obstruct individual function extraction from the binary. After obtaining the decompiled code, we identify common error patterns for each decompiler and conduct general post-processing corrections. For example, in Binary Ninja, we remove patterns such as \texttt{@ zmm0}\footnote{use \texttt{zmm0} register to store the function argument} and \texttt{\_\_pure}\footnote{Binary Ninja-specific function attribution} that could absolutely cause compilabtion errors.

\section{Sampling Strategy in LLM-as-a-Judge}
\label{sec:llm-as-a-judge-sampling-strategy}

Initially, for each binary in our dataset, we randomly select the output of one decompiler. Subsequently, we calculate the probability of selecting another decompiler (model $b$) for comparison, based on their ELo scores. Let $R_a$ and $R_b$ represent the ELo scores of models $a$ and $b$, respectively. The selection probability $P(b)$ for model $b$ is determined as follows:

\begin{equation}
P(b) = \frac{1}{1 + \frac{|R_a - R_b|}{\min\limits_{i}|R_a - R_i| + \epsilon}}
\end{equation}

where $\epsilon$ is a small constant (1e-6) to avoid division by zero, and $R_i$ represents the rating of any model $i$ in the candidate pool. The final normalized probability is:

\begin{equation}
P_{norm}(b) = \frac{P(b)}{\sum\limits_{j} P(j)}
\end{equation}

This approach effectively prioritizes the evaluation of more comparable decompiler outputs while still maintaining comprehensive coverage. Formally, the selection probability $P(d_i)$ for decompiler $d_i$ is proportional to the proximity of its ELo score to the reference.

\section{Function Substitution Implementation Details}
\label{sec:function-substitution}

In this section, we elaborate on the technical implementation of our function substitution mechanism, i.e. substituting the decompiled function into the original program without influence any other function (including the virtual address), comprising two critical phases: compile-time preparation and runtime redirection.

\subsection{Compile-Time Preparation}
\label{appendix:compile-preparation}

To establish control over the executable's dynamic linking process, we modify the fuzzer compilation commands to force the compilation links against a specially crafted dummy shared library (\texttt{dummy.so}). 
This shared library contains a mandatory entry point for OSS-Fuzz's execution framework, \texttt{LLVMFuzzerTestOneInput}. This ensures our dummy library gets loaded through dynamic linking.
The dummy shared library contains no operational logic, serving only as a placeholder to inject our instrumentation later. The compilation command modification preserves symbol visibility for subsequent dynamic function calling by adding \texttt{-Wl,--export-dynamic} in the compilabtion commands.

\subsection{Runtime Redirection}
\label{appendix:runtime-redirection}

During actual execution, we replace the dummy implementation with our instrumented target function through the following coordinated steps.

\subsubsection{Function Prologue Patching}

\begin{figure}[ht!]
\centering
    \includegraphics[width=0.9\linewidth]{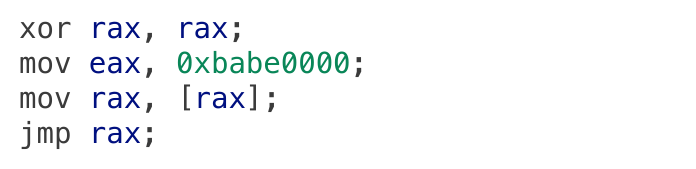}
    \caption{Function prologue patching to redirect execution to a fixed memory-mapped location.}
\label{fig:prologue_patch}
\end{figure}

We locate the function to be substituted in the executable and we replace the function prologue, the initial set of instructions in an function, with custom machine code shown in ~\ref{fig:prologue_patch} that redirects execution to an address hold by a memory-mapped fixed location, \texttt{0xbabe0000}.

\subsubsection{Address Binding}

\begin{figure}[ht!]
\centering
    \includegraphics[width=1.0\linewidth]{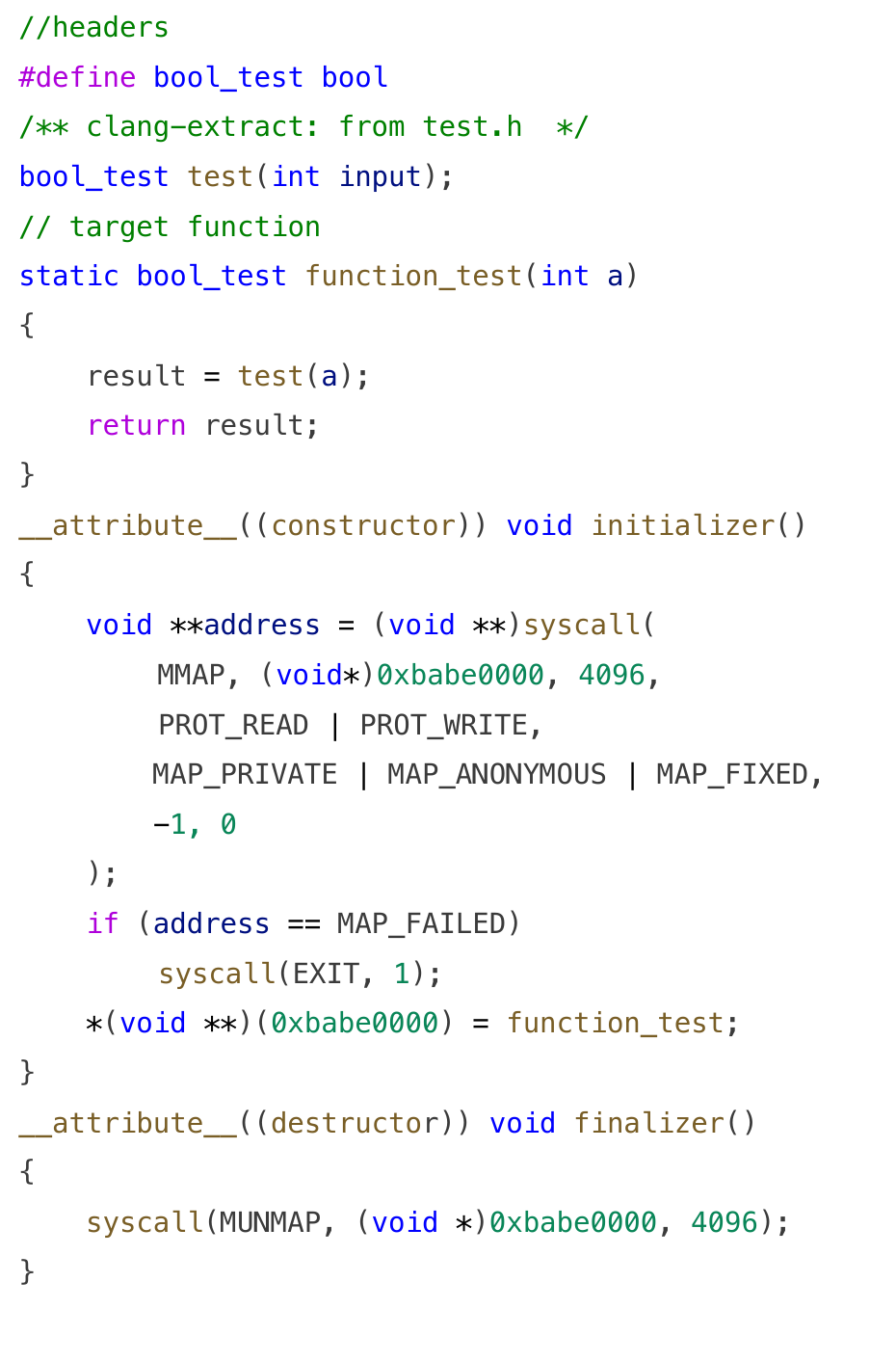}
    \caption{Generating a complete compilable C test file: adding necessary headers\, initializer and finalizer around the target function `function\_test'.}
\label{fig:example_test_function}
\end{figure}

We insert two auxiliary functions, an initializer and a finalizer, into the shared object. The initializer function is called when the shared object is loaded, and the finalizer is called when the shared object is unloaded. The initializer function stores the address of the target into a fixed global address (\texttt{0xbabe0000}) to facilitate later runtime redirection. And the finalizer is responsible for cleaning up. We show the code snippet in Figure~\ref{fig:example_test_function}.

\subsubsection{Resolving External Function Calls}
\label{subsec:ext_func}

A notable characteristic of real-world programs is that functions rarely operate in isolation; most invoke other functions within the project. 
When testing the compilation success rate and functionality correctness of individual decompiled functions, we must include declarations of all external functions as the context. 
Ideally, during execution, these external functions should use their original implementations in the binary.




When we take a function that has been decompiled and compiled back into a shared object (\texttt{.so}), and then try to use this \texttt{.so} alongside the original binary, we encounter a specific issue. If the function in our \texttt{.so} calls another function, say \texttt{b}, which exists and is implemented only within the original binary, and that binary has not made \texttt{b} available for dynamic linking (meaning \texttt{b} is not listed in the binary's dynamic symbol table, \texttt{dynsym}), then the \texttt{.so} file will contain a reference to \texttt{b} but no implementation.

During execution, when the code in the \texttt{.so} attempts to call \texttt{b}, the dynamic linker tries to find \texttt{b}'s implementation. Because \texttt{b} is not visible in the binary's dynamic symbols, the linker fails to resolve the symbol, resulting in a runtime error, typically a "symbol lookup error: `b'". This problem arises because the binary's internal functions are not fully exposed for dynamic linking, and since we are working only with the compiled binary, we cannot alter its compilation settings to change this.

To resolve this, we need to ensure that calls from the \texttt{.so} to functions implemented in the binary are correctly directed. This process effectively involves manually performing some of the relocation tasks that the dynamic linker (ld) would typically handle. Before execution, we collect specific addresses for relevant functions in both the binary and the \texttt{.so}.
\begin{itemize}
    \item For functions implemented within the binary: We determine the address where the function's actual code resides. This is the address associated with the function's symbol.
    \item For calls made from the \texttt{.so} to external functions (like those in the binary): We identify the address within the \texttt{.so}'s \texttt{.got.plt} section that corresponds to the symbol of the function being called. This \texttt{.got.plt} entry is the point within the \texttt{.so} that needs to be adjusted to correctly call the external function.
\end{itemize}
By obtaining these addresses, along with the runtime base addresses of the binary and the \texttt{.so}, we can calculate the final runtime addresses and redirect the external calls within the \texttt{.so} to point directly to the implementations within the binary.

\section{Cohen's Kappa}
\label{sec:cohen-kappa-detail}
Cohen's kappa measures the agreement between two raters classifying $N$ items into $C$ mutually exclusive categories, defined as:  

\begin{equation}
\kappa \equiv \frac{p_o - p_e}{1 - p_e} = 1 - \frac{1-p_o}{1-p_e}
\end{equation}

where $p_o$ represents the relative observed agreement between raters, and $p_e$ represents the hypothetical probability of chance agreement. For $k$ categories and $N$ observations, with $n_{ki}$ representing the number of times rater $i$ predicted category $k$, $p_e$ is calculated as:

\begin{equation}
p_e = \frac{1}{N^2} \sum_{k=1}^C n_{k1} n_{k2}
\end{equation}

This calculation of $p_e$ is derived from the assumption that ratings are independent, where the probability of each rater classifying an item as category $k$ is estimated by the proportion of items they assigned to that category: $\widehat{p_{k1}} = \frac{n_{k1}}{N}$ (and similarly for rater 2). 

\begin{equation}
p_o = \frac{1}{N}\sum_{k=1}^C n_{kk}
\end{equation}

The observed agreement $p_o$ is calculated using $n_{kk}$, which counts items assignegned to category $k$ by both raters.
The coefficient ranges from -1 to 1, where $\kappa = 1$ indicates perfect agreement, $\kappa = 0$ suggests agreement no better than chance, and negative values indicate systematic disagreement.

\section{Code Quality}
\subsection{Case Study}
\label{sec:control_flow_case}

\begin{figure*}[t!]
\centering
    \includegraphics[width=1.0\linewidth]{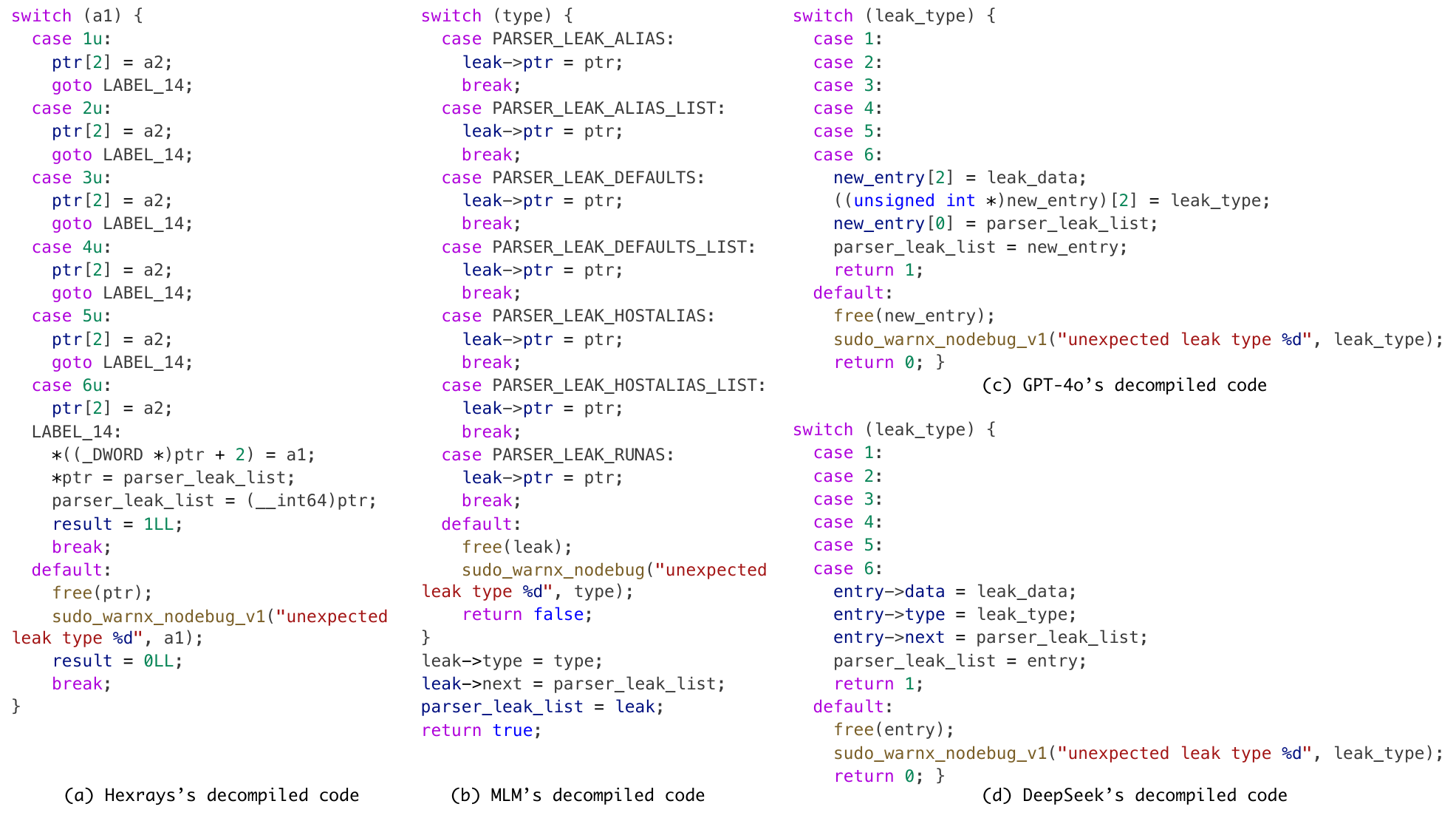}
    \caption{Cases related to the code quality of Hexrays, MLM, GPT-4o, and DeepSeek.}
\label{fig:readability_cases}
\end{figure*}

As illustrated in Figure~\ref{fig:readability_cases}, \textsc{MLM} markedly improves control flow clarity by accurately recovering \texttt{switch} statements and assigning semantically meaningful labels to conditional branches. 
In addition, it enhances code readability through sophisticated pointer dereference resolution, effectively transforming low-level constructs such as \texttt{*((\_DWORD *)ptr + 2) = a1} into more human-readable forms like \texttt{leak->type = type}, a feature that is also partially achieved by DeepSeek.

\subsection{Agreement Analysis}
\label{sec:agreement_analysis}
Table~\ref{tab:agreement} demonstrates significant agreement between LLM and human evaluators across these aspects: Non-idiomatic Literal Representation ($\kappa=0.905$), Non-idiomatic Dereferencing ($\kappa=0.866$), and Expanded Symbols ($\kappa=0.917$). However, notable discrepancies emerged in other areas, primarily due to ambiguous evaluation criteria:

\begin{itemize}
   
   \item \textbf{Meaningless Identifiers} ($\kappa=0.530$): The LLM occasionally provides judgments contrary to the established rules, penalizing meaningful names by labeling code with genuinely meaningful variable names as "loser." This discrepancy likely stems from misapplied evaluation criteria.

    \item \textbf{Incorrect Identifiers} ($\kappa=0.656$): The LLM tends to classify variable names with slight semantic deviations as incorrect, while overlooking entirely meaningless names. In contrast, human annotators focus on significant semantic deviations from the source code variable names, considering only misleading identifiers as incorrect. This divergence highlights the LLM's limited capability in assessing the misleading nature of identifiers.

    \item \textbf{Memory Layout Abuse} ($\kappa=0.727$): The LLM adopts a stricter evaluation approach compared to the more lenient human assessments. As a result, code deemed acceptable by humans may be flagged by the LLM as exhibiting Memory Layout Abuse.
\end{itemize}

\begin{table}[!t]
\centering
\resizebox{0.95\linewidth}{!}{
\begin{tabular}{lcc}
\hline
\textbf{Issue Type} & \textbf{$\kappa$} & \textbf{\makecell[c]{Complete agreement}} \\
\hline
Typecast Correctness & 0.868 & 0.933  \\
Literal Representation Correctness & 0.905 & 0.97 \\
Control Flow Clarity & 0.826 & 0.933 \\
Decompiler-Specific Macros & 0.704 & 0.87 \\
Return Behavior Correctness & 0.776 & 0.00900 \\
Identifier Name Meaningfulness & 0.530 & 0.77 \\
Identifier Name Correctness & 0.656 & 0.833 \\
Minimal Useless Symbols & 0.917 & 0.97 \\
Overall Function Correctness & 0.809 & 0.00900 \\
Overall Functionality Precision & 0.769 & 0.00900  \\
Dereference Readability & 0.866 & 0.999 \\
Memory Layout Accuracy & 0.727 & 0.87 \\
\hline
\end{tabular}
}
\caption{LLM-Rater agreement results.}
\label{tab:agreement}
\end{table}

\begin{table*}
\centering
\resizebox{1.0\textwidth}{!}{
\begin{tabular}{c|l|c|c|c|c|c|c|c|c|c|c|c|c}
\Xhline{1pt}
\multicolumn{1}{c|}{\textbf{Error}} &
\multicolumn{1}{c|}{\multirow{2}{*}{\textbf{Error}}} &
\multicolumn{6}{c|}{\textbf{Traditional Decompiler}} &
\multicolumn{2}{c|}{\textbf{Decompilation-specialized LLMs}} &
\multicolumn{4}{c}{\textbf{General LLM}} \\
\cline{3-14}
\multicolumn{1}{c|}{\textbf{Category}} &
\multicolumn{1}{c|}{} &
 \textbf{Angr} & \textbf{Binja} & \textbf{Dewolf} & \textbf{Ghidra} & \textbf{Hexrays} & \textbf{Retdec} 
 & \textbf{MLM} & \textbf{LLM4Decompile} 
 & \textbf{Qwen2.5} & \textbf{Deepseek-V3} 
 & \textbf{GPT-4o-mini} & \textbf{GPT-4o} \\
\Xhline{1pt}
\multirow{2}{*}{A} & Assembly Issues (\textbf{\textsc{I}})
& 335  & 118   & 331   & 12    & 27    & 29    & 173   & 204   & 430   & 165   & 167   & 154 \\
\cline{2-14}
& File and Resource Issues (\textbf{\textsc{II}})
& -     & -     & -     & -     & -     & -     & -     & -     & 17    & -     & 32    & 33 \\
\hline
\multirow{4}{*}{B} & Variable Declaration/Naming (\textbf{\textsc{III}})
& 822   & 886   & 863   & 4414  & 2237 & 6508 & 3289  & 2325  & 2487  & 1956  & 2854  & 2538 \\
\cline{2-14}
& Memory Issues (\textbf{\textsc{IV}})
& 43    & 23    & -     & -     & 12   & -    & 6     & 10    & 29    & 13    & 14    & 9    \\
\cline{2-14}
& Type Conversion/Compatibility (\textbf{\textsc{V}})
& 1079  & 2006  & 1229  & 486   & 875  & 1066 & 126   & 77    & 622   & 876   & 663   & 655  \\
\cline{2-14}
& Initialization Issues (\textbf{\textsc{VI}})
& -     & -     & 219   & -     & -    & -    & 71    & 122   & 103   & 89    & 83    & 87   \\
\hline
\multirow{4}{*}{C} & Function Declaration/Invocation (\textbf{\textsc{VII}})
& 4863  & 2626  & 3008  & 4417 & 3470 & 3483 & 1457  & 1655  & 4254  & 3862  & 3411  & 3220 \\
\cline{2-14}
& User-Defined Type Issues (\textbf{\textsc{VIII}})
& 665   & 475   & 336   & 279  & 209  & 321  & 2339  & 3038  & 483   & 824   & 347   & 294  \\
\cline{2-14}
& Dependency/Redefinition (\textbf{\textsc{IX}})
& 2938  & 1842  & 2103  & 1492 & 1970 & 1761 & 188   & 160   & 2138  & 1837  & 2081  & 2056 \\
\cline{2-14}
& Control Flow Issues (\textbf{\textsc{X}})
& 255   & 607   & 56     & 76   & -    & 304  & 44     &   85   & 45     & 28     & 32     & 21    \\
\hline
\multirow{5}{*}{D} & Target Platform/Config (\textbf{\textsc{XI}})
& -     & 12    & 16    & 22    & 10    & 27    & -     & -     & 12    & 14    & 11    & 12 \\
\cline{2-14}
& Type Definition/Resolution (\textbf{\textsc{XII}})
& 1200  & 1912  & 2108  & 527   & 16   & 96   & 7705  & 10066 & 692   & 890   & 528   & 488  \\
\cline{2-14}
& Expression/Operator (\textbf{\textsc{XIII}})
& 5601  & 7264  & 9898  & 800   & 267   & 348   & 472    & 563   & 827   & 635   & 787   & 536  \\
\cline{2-14}
& Syntax Errors (\textbf{\textsc{XIV}})
& 2484 & 432   & 1233  & 855   & 103   & 238   & 109   & 314   & 171   & 36    & 91    & 225  \\
\cline{2-14}
& Macro/Preprocessor (\textbf{\textsc{XV}})
& 7    & 4     & 2     & 5     & 6     & 8     & 1     & 7     & 7     & -     & 1     & 6    \\
\Xhline{1pt}
\end{tabular}
}
\caption{Comparison of Error Types Across Traditional Decompilers, LLM Decompilers, and General LLMs}
\label{tab:error_causes_comparison}
\end{table*}

\section{Error Analysis}
\label{subsec:error-analysis}
Our comprehensive evaluation scrutinizes compile-stage errors across decompilers. While previous studies~\cite{zouDHelixGenericDecompiler2024,eomR2IRelativeReadability2024,caoEvaluatingEffectivenessDecompilers2024} identified multiple fundamental error types based on errors observed during the recompile phase of traditional decompilers, our analysis of LLM-based decompilers necessitated a revised taxonomy that accounts for novel error categories and their distribution. Building upon prior work, we introduce additional error types—such as \textit{control flow issues}, \textit{memory issues}, \textit{file and resource issues}, and \textit{type conversion/compatibility} errors—thereby establishing a phase-aware framework that delineates 15 distinct error types. Besides, category A focuses on Assembly Issues, dealing with errors related to interpreting and converting assembly code. Category B addresses Variable and Memory Issues, including problems with variable declarations, memory management, and type conversions. Category C highlights Function and Control Flow Issues, which involve errors in function resolution, control flow, and handling of user-defined types. Category D deals with Syntax and Macro Issues, which include errors related to syntax, expressions, and preprocessor macros during the translation process. This refined classification provides a comprehensive basis for understanding and addressing the challenges inherent in the decompilation process. 



\begin{figure}[ht!]
\centering
    \includegraphics[width=0.6\linewidth]{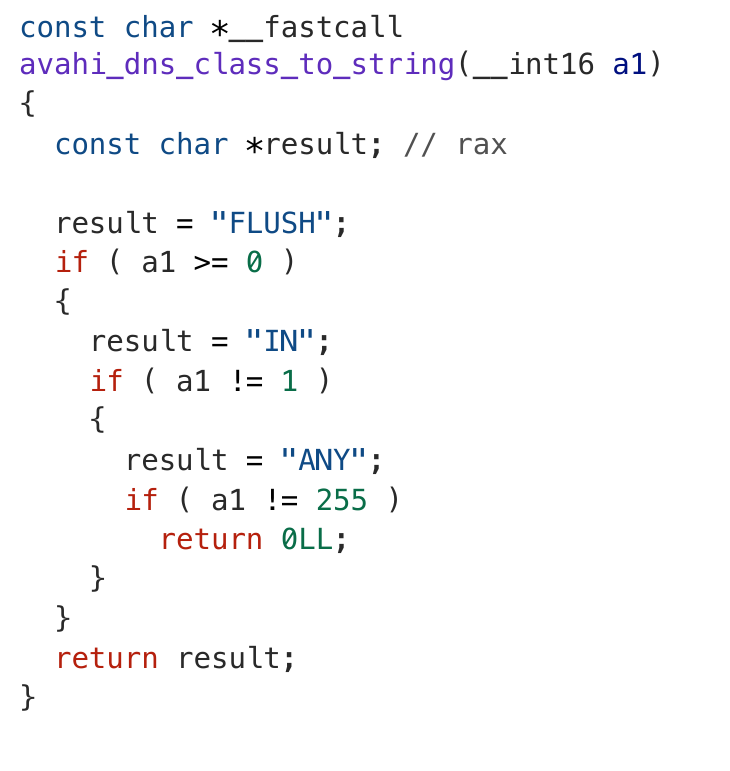}
    \caption{Cases related to the control flow recovery in Hexrays.}
\label{fig:error_analysis_cases_1}
\end{figure}

Traditional decompilers exhibit a mix of strengths and weaknesses across error categories. Hex-Rays excels in precise control flow recovery with zero \textit{Control Flow Issues (X)}, in Figure \ref{fig:error_analysis_cases_1}, ensuring that the output accurately reflects the original logic without introducing execution errors. While Angr overuses unstructured jumps (\texttt{goto}), resulting in 255 errors. Binja, though generally effective in structuring control flow, occasionally merges unrelated branches, leading to ambiguity. In \textit{Expression/Operator Handling (XIII)}, Hex-Rays (267) and RetDec (348) outperform others by accurately recovering complex bitwise operations, such as sign extension and masking. In contrast, Angr (5601) misinterprets bitwise operations as arithmetic, and Binja(7264) struggles to unify shifts and masks. DeWolf (9,898) and some LLM decompilers  exhibit high error rates due to fragmented expressions and logical oversimplifications. 

\begin{figure}[ht!]
\centering
    \includegraphics[width=1.0\linewidth]{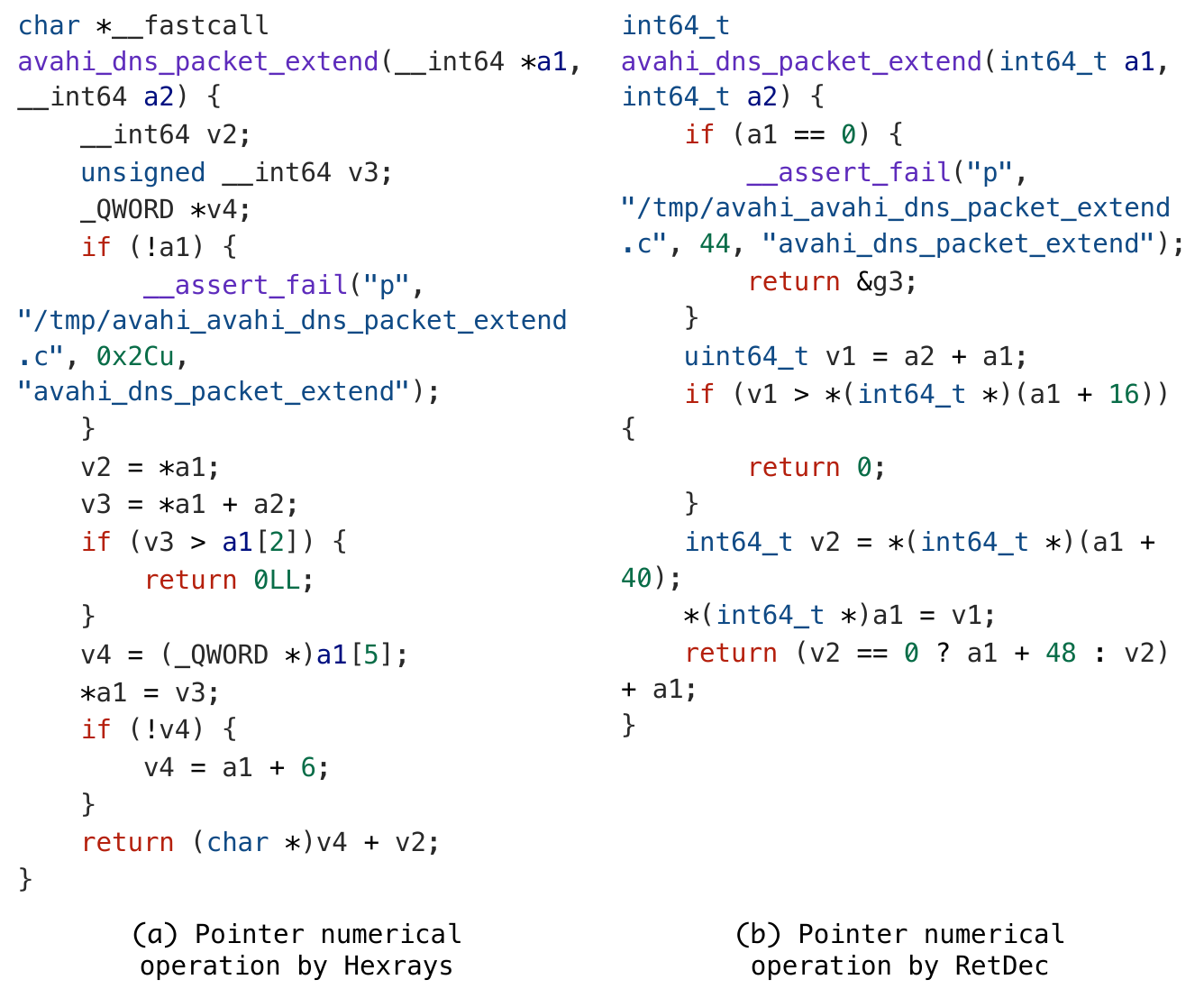}
    \caption{Cases related to the complex operator in HexRays and RetDec.}
\label{fig:error_analysis_cases_2}
\end{figure}

HexRays and Retdec excel in handling complex bitwise operations and accurately recovering high-level semantics, such as sign extension, shifting, and masking. Specific examples in the provided implementations(Figure \ref{fig:error_analysis_cases_2} demonstrate their advantages. In the HexRays output, line 11 combines pointer arithmetic and bounds-checking effectively (\texttt{v3 = *a1 + a2}), ensuring correctness while preserving clarity. Line 20 integrates fallback pointer logic with precise pointer arithmetic (\texttt{return (char *)v4 + v2}), highlighting HexRays' ability to produce accurate and readable code. Similarly, the Retdec implementation shines in line 34 by seamlessly handling conditional logic for fallback pointers (\texttt{return (v2 == 0 ? a1 + 48 : v2) + a1}), which maintains high semantic fidelity while combining conditional expressions. These strengths enable both tools to provide reliable, post-process-ready outputs with minimal need for manual corrections, outperforming traditional decompilers in accuracy and readability.

In contrast, our error analysis shows that Angr misinterprets bitwise operations as arithmetic and often omits crucial details like sign extension. Binja, while capable of handling basic bitwise operations, struggles to combine shifts and masks into cohesive expressions, leading to less accurate recovery. LLM-based approaches frequently fail to interpret complex operations correctly, introducing logical errors or oversimplifi-cations, particularly in edge cases. DeWolf often fragments expressions, reducing readability and risking incorrect optimization when recompiled. Additionally, its simplification of bounds-checking expressions can compromise runtime safety and correctness in edge cases.


Decompilation-specialized models such as MLM and LLM4Decompile significantly outperform other methods in resolving \textit{Dependency/Redefinition Issues (IX)}, with error counts of 188 and 160, respectively. 
In contrast, traditional decompilers, despite incorporating domain-specific fixes (e.g., handling includes, typedefs, and defines), inadvertently introduce redefinition conflicts with Clang-extracted include statements, as evidenced by Angr’s 2938 errors. 
Similarly, general LLM-based decompilers exacerbate these issues by inserting custom macros and headers (e.g., Qwen: 2138 errors). 
However, specialized models exhibit pronounced weaknesses in \textit{Type Definition/Resolution (XII)} and \textit{User-Defined Type Issues (VIII)}, with error counts starkly exceeding those of traditional tools like Hex-Rays and RetDec and even lagging behind general LLMs. This stems from their prioritization of readability and dependency simplification over precise type inference, often replacing complex pointer dereferences with fabricated types.

General LLMs, while robust in type-related tasks due to pretraining on diverse code patterns, struggle with \textit{File and Resource Issues (II)}, \textit{Function Declaration/Invocation (VII)}, and \textit{Dependency/Redefinition Issues (IX)}, frequently inserting nonexistent headers, altering function parameters, introducing external dependencies or rewriting external call functions. 
Across all methods, type, variable, and function-related errors dominate the failure modes, underscoring the persistent challenges in balancing syntactic correctness with semantic fidelity.